# Large Transport Critical Current Densities of Ag Sheathed (Ba,K)$Fe_2As_2$+Ag Superconducting Wires Fabricated by an *ex-situ* Powder-in-Tube (PIT) Process

**Kazumasa Togano, Akiyoshi Matsumoto, and Hiroaki Kumakura**

National Institute for Materials Science, Tsukuba, Ibaraki 305-0047, Japan

## Abstract

We report large transport critical current densities observed in Ag-added (Ba,K)$Fe_2As_2$ superconducting wires prepared by an *ex-situ* powder-in-tube (PIT) process. The wire has a simple composite structure sheathed only by Ag. A precursor bulk material prepared by a melting process was ground into powder and put into a Ag tube. The composite was then cold worked into a wire and heat treated at 850ºC for sintering. Transport critical current densities, $J_c$, at 4.2 K are $1.0\times10^4$ A/cm$^2$ ($I_c$ = 60.7 A) in self field and $1.1\times10^3$ A/cm$^2$ ($I_c$ = 6.6 A) in 10 T. These are the highest values reported so far for the Fe-based superconducting wires.

The discovery of superconductivity in Fe-based superconductors[1)] has elicited enormous excitement in the fields of both superconductor applications and condensed matter physics. Up to now, many families of Fe based superconductors, namely, REOFeAs(1111),[2)] LiFeAs(111),[3)] $BaFe_2As_2$(122),[4)] FeSe(11),[5)] and the pnictides with perovskite-type blocking layer[6,7)] have been discovered, which show high-temperature superconductivity by appropriate doping. In addition to the high transition temperature, $T_c$, these Fe-based superconductors were reported to have a very high upper critical field, $H_{c2}$, bringing the hope of high field applications as wires and bulks.[8-10)] Relatively smaller anisotropy of 11 and 122 compounds than those of cuprate superconductors is also attractive for magnet applications because it is expected to bring about a higher irreversibility field, $H_{\mathrm{irr}}$.[10)] In order to evaluate the potentiality for wire applications, the development of wire processing technique is essential. The first attempt of wire fabrication was carried out by the group of Institute of Electrical Engineering (IEE), Chinese Academy of Science, by applying a powder-in-tube (PIT) process for LaFeAsOF,[11)] SmFeAsOF,[12)] and (Sr,K)$Fe_2As_2$.[13)] The PIT process has been the most conventional process for the fabrication of BiSrCaCuO and $MgB_2$ superconduting wires. A similar PIT process was also applied to Fe(Se,Te) by Mizuguchi *et al.*[14)] They succeeded in measuring the transport critical current density, $J_c$, although it was as small as the order of 1 A/cm$^2$ in applied magnetic fields. Immediately after that, the IEE group succeeded in observing large transport critical currents, $I_c$, at 4.2 K for the PIT-processed wires of (Sr,K)$Fe_2As_2$[15,16)] and SmFeAsOF.[17)] They reported the transport $J_c$ of 3750 A/cm$^2$ ($I_c$ = 37.5 A) in self-field and at 4.2 K for an *ex-situ* processed $Sr_{0.6}K_{0.4}Fe_2As_2$+Pb wire with an Fe/Ag double sheath.[16)]

In this paper, we report the further improvement of the transport critical current properties of the *ex-situ*-PIT-processed 122 wires, which was achieved by using a precursor material of (Ba,K)$Fe_2As_2$ with Ag addition prepared by the melting process at a high temperature. The wire has a simple composite structure sheathed only by Ag. This will be advantageous for the production of practical wires, because the pure Ag has good workability sufficient for scale-up production and high electric conductivity suitable for conductor stabilization.

The starting materials are pieces of a few mm size of Ba(99%), K(98%), commercially available FeAs alloy(99.5%), and Ag(99.995%) . The use of pieces instead of powder makes the handling of the materials much easier even in a glove box.

These materials with a nominal composition of Ba:K:FeAs:Ag = (0.6×1.1):(0.4×1.2):2:0.5 were put into a BN crucible. The Ag was added in order to improve the grain connectivity, as reported for the (Sr,K)$Fe_2As_2$ system.[16,18,19] In order to compensate for the loss of Ba and K elements during high-temperature heat treatment, 10% extra Ba and 20% extra K were added to the stoichiometric composition of $(Ba_{0.6}K_{0.4})Fe_2As_2$. The BN crucible was then put into a stainless steel pipe. Both ends of the stainless steel pipe were pressed and welded by arc melting in an argon atmosphere. In order to have a good mixing of constituent elements, the heat treatment was carried out at a high temperature above the melting point of the FeAs compound (~1050ºC).[20] The sealed stainless pipe was put into a box furnace held at 1100ºC for 5 min and then the furnace was cooled by switching off the current source.

Figure 1 shows the powder X-ray diffraction (XRD) pattern of the obtained bulk material. Despite starting from the pieces, the diffraction pattern has strong peaks of 122 phase indicating that the reaction to form 122 phase completes in a short time in the molten state. The diffraction pattern also includes peaks of the added Ag and small peaks of impurity phases such as the FeAs compound. The good formation of 122 superconducting phase was also confirmed by the strong signal of diamagnetism, which is shown in the inset of Fig. 1. The onset $T_c$ is ~35 K, which is lower than the ideal $T_c$ of 122 phase (38 K).[4] The origin of the lower $T_c$ is not clear, however, it is possible that the added Ag partially substitutes in 122 phase during the high-temperature heat treatment.

We used the obtained bulk sample with Ag addition as a precursor for the fabrication of (Ba,K)$Fe_2As_2$ wires by an *ex-situ* PIT process. The bulk sample was crushed and ground into powder form in an agate mortar and put into a Ag pipe of 6 mm outer diameter and 4 mm inner diameter. The composite was then cold worked into a wire of 2.0 mm diameter by grooved rolling and finally swaging. A short sample of 35 mm in length cut from the wire was put into a stainless steel pipe together with small pieces of Ba, K, and FeAs alloy in order to suppress the loss of the components from both ends of the wire during the heat treatment. Both ends of the stainless steel pipe were pressed and sealed by arc welding in an Ar atmosphere. The sealed stainless steel pipe was then heat treated at 850ºC for 3, 15, and 30 h and cooled to room temperature in the furnace.

Figure 2(a) shows the transverse cross section of the wire heat treated at 850ºC

for 30 h observed using an optical microscope. No appreciable reaction was observed at the boundary between the core and the Ag sheath, indicating good suitability of Ag as a sheath material for the fabrication of PIT wires. Figure 2(b) is a higher magnification optical photograph observed inside the core, which shows that the core is composed of grey matrix of fine grains, dispersed white particles, and dark spots of void probably produced by the sintering process. Taking account of the results of XRD and magnetization measurement shown in Fig. 3, we conclude that the matrix and the white particles are 122 phase and Ag, respectively. The onset $T_c$ is ~35 K, which is almost same with that of the precursor. A detailed study on composition and grain structure is now in progress.

We carried out transport critical current measurements in liquid helium and applied magnetic fields using 12 and 18 T superconducting magnets. Voltage and current leads were attached to the surface of the Ag sheath by soldering. The distance between the voltage taps is 10 mm. A magnetic field was applied perpendicularly to the wire axis. Figure 4 shows typical curves of the voltage as a function of applied current measured in an 18 T superconducting magnet for the wire heat treated at 850°C for 30 h . The curves show a clear transition from the zero resistance state to the resistive state, indicating that the boundary between the superconducting core and sheath has good electrical contact and, hence, good current transfer. Critical current, $I_c$, is defined as the current of 1 μV voltage appearance. The $I_c$ at 4.2 K is as high as 60.7 A in self-field and 5.2 A in 18 T, which is the highest magnetic field in this work.

Transport critical current density, $J_c$, was estimated by dividing the $I_c$ by the cross-sectional area of the superconducting core. Figure 5 shows the plots of transport $J_c$ as a function of applied magnetic field for the wires heat treated at 850°C for 3, 15, and 30 h, respectively. The 3 h sample shows extremely low transport $J_c$ of less than $10^2$ A/cm$^2$ in applied magnetic fields, probably due to imperfect grain connection. However, the wires heat treated for 15 and 30 h show much improved transport $J_c$. For the 15 h sample, the measurement was initially carried out using a 12 T superconducting magnet, and then an 18 T superconducting magnet for the same sample after a week. The curves almost coincide as shown in Fig. 5, indicating that no appreciable degradation occurred during this period. The wire heat treated for 30 h shows the highest transport $J_c$ in this work. The transport $J_c$ in self-field is $1.01\times10^4$ A/cm$^2$ ($I_c$ = 60.7 A), which is the highest value reported so far for Fe-based superconducting wires. In applied magnetic fields, $J_c$ rapidly drops about one order of magnitude and then shows a very small field dependence as reported for Fe(Se,Te)[14] and (Sr,K)$Fe_2As_2$ [16] wires. The transport $J_c$ at 10 T is 1.1×10$^3$ A/cm$^2$, which is also the highest value of the Fe-based superconducting wires. We expect that the transport $J_c$ can be more increased by reducing the amounts of porosity and inclusions. In Fig. 5, it is also noticed that the curves show a very small field dependence up to the strongest magnetic field of 18 T. This is consistent with the high values of the upper critical field $H_{c2}$ reported for (Ba,K)$Fe_2As_2$ single crystal .[10]

In summary, we fabricated Ag-sheathed (Ba,K)$Fe_2As_2$+Ag superconducting wires with large transport $J_c$ by an *ex-situ* PIT process. The precursor material was prepared by a high-temperature synthesis above the melting point, which was then ground into powder form. The powder was then put into a Ag pipe and the composite was cold worked into a wire and heat treated at 850ºC for sintering. The transport $J_c$ (at 4.2 K) values obtained are 1.0×10$^4$ A/cm$^2$ ($I_c$ = 60.7A), 1.1×10$^3$ A/cm$^2$ ($I_c$ = 6.6 A), and 8.7×10$^2$ A/cm$^2$ ($I_c$ = 5.2 A) in applied magnetic fields of 0, 10, and 18 T, respectively. Good grain connectivity in the precursor particles and good interparticle connection obtained by the Ag addition are considered to be responsible for the enhanced transport $J_c$. We believe that further improvement of transport critical current properties is possible by optimizing the processing conditions.

**Acknowlegdements**

This work was supported by a Grant-in Aid for “Transformative Research project on Iron Pnictides (TRIP)” from the Japan Science and Technology Agency (JST) and the Japan Society for the Promotion of Science (JSPS) through its “Funding Program for World-Leading Innovative R&D on Science and Technology (FIRST) Program. We acknowledge Dr. H. Fujii and Mr. S. J. Ye of the National Institute for Materials Science for their assistance in $I_c$ measurement and Prof. Y. W. Ma of the Chinese Academy of Science, Beijing, for his useful discussions.

**Referfences**

**Figure Captions**

Fig. 1. Powder X-ray diffraction pattern of the precursor material of (Ba,K)$Fe_2As_2$+Ag prepared by a high-temperature synthesis. The inset shows the magnetization vs. temperature curve measured for the bulk.

Fig. 2. (a) Transverse cross section observed by an optical microscope for the wire heat treated at 850ºC for 30 h.
(b) Optical microstructure of the core observed on the polished cross section of the wire heat treated at 850ºC for 30 h. .

Fig. 3. Powder X-ray diffraction pattern of the wire heat treated at 850ºC for 30 h. Powder was scraped off from the Ag sheath. The inset is the magnetization vs temperature curve. Measurement was done for a 2 mm length wire as Ag sheathed.

Fig. 4. Typical voltage vs applied current curves measured for the wire heat treated at 850ºC for 30 h. Measurement was carried out in liquid helium (4.2 K) using an 18 T superconducting magnet.

Fig. 5. Transport $J_c$ as a function of applied magnetic field of the wires heat treated at 850ºC for 3, 15, and 30 h. The measurement was carried out in liquid helium (4.2 K) using 12 and 18 T superconducting magnets.

Fig. 1

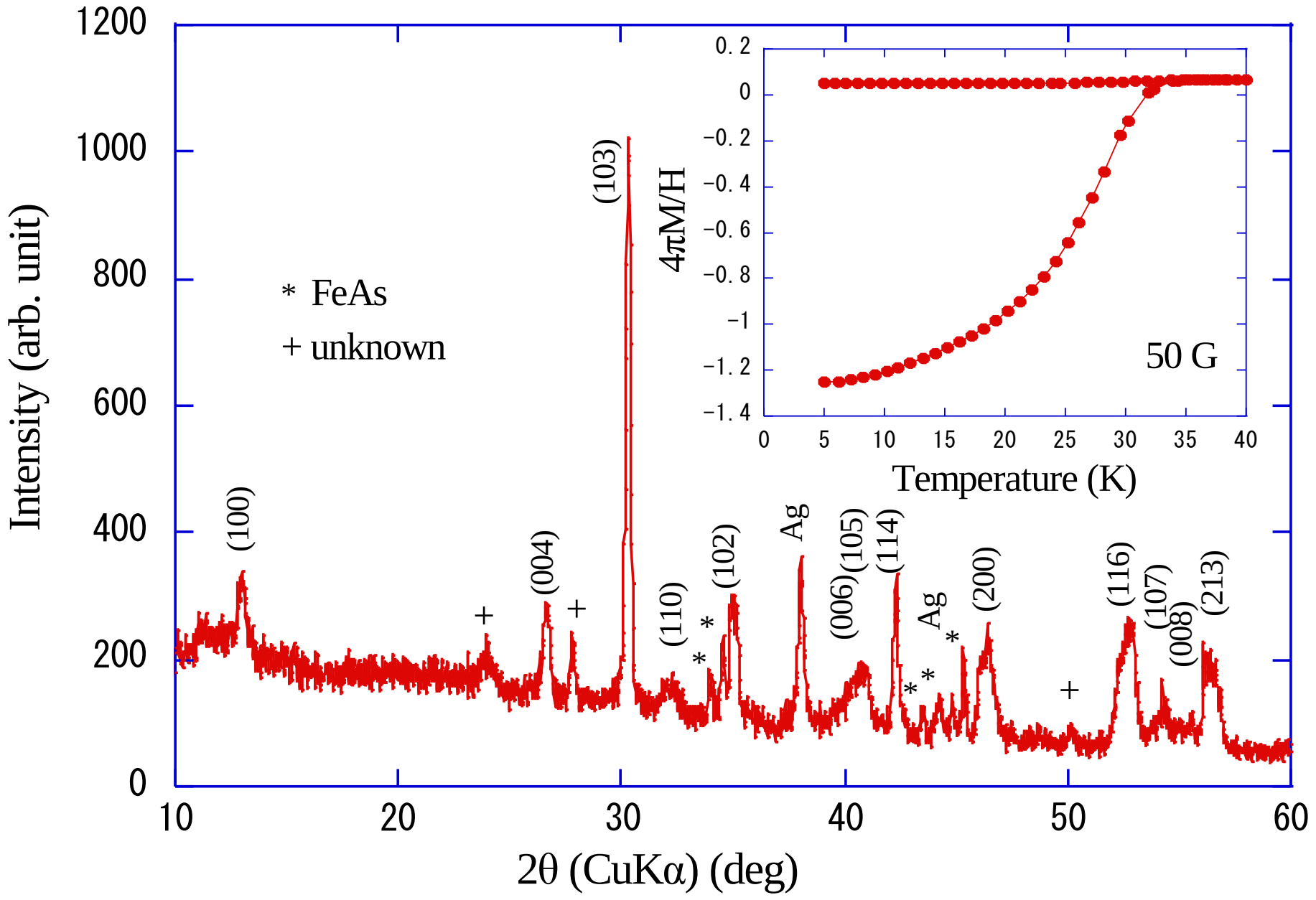

Fig. 2

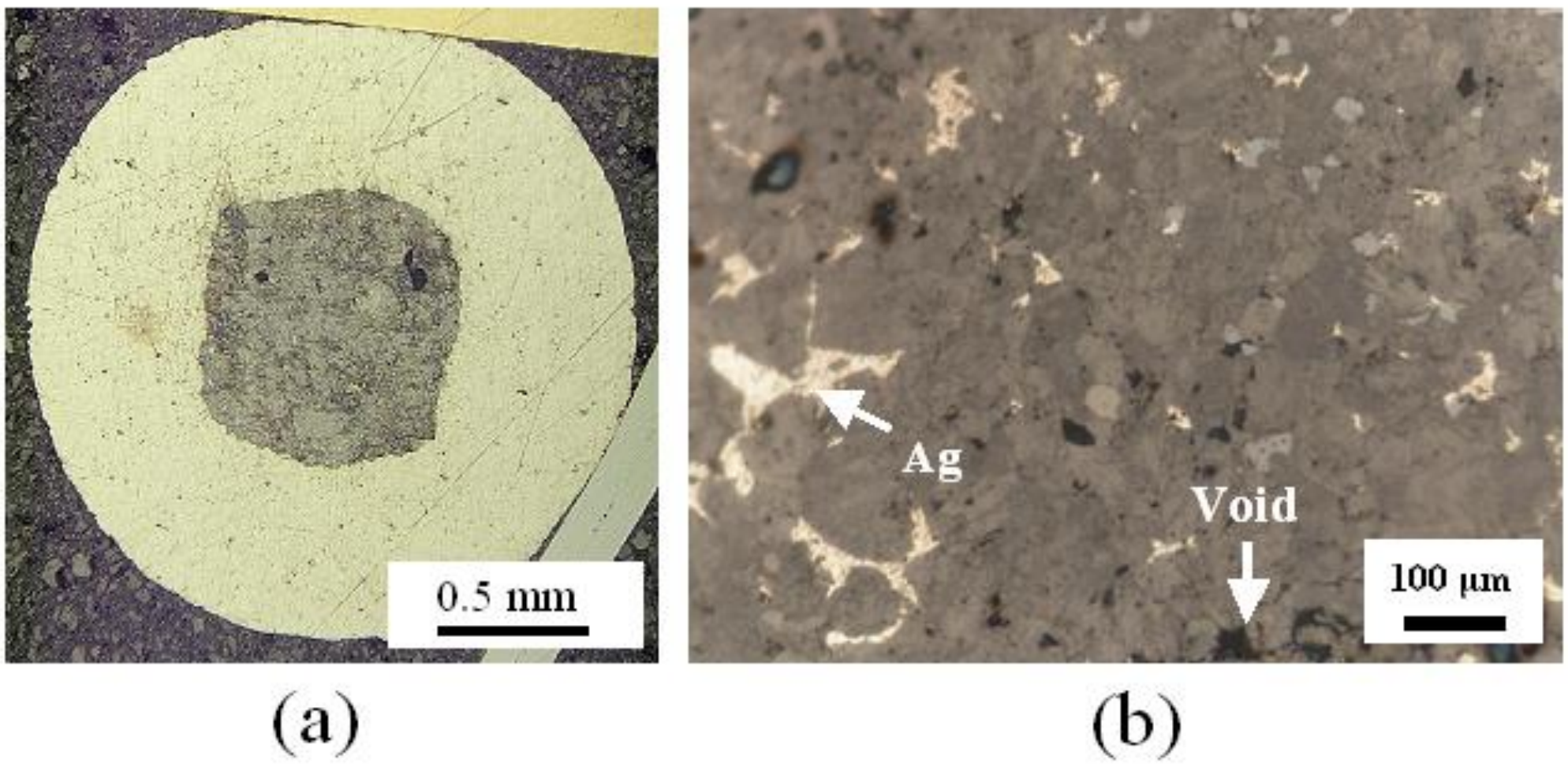

Fig. 3

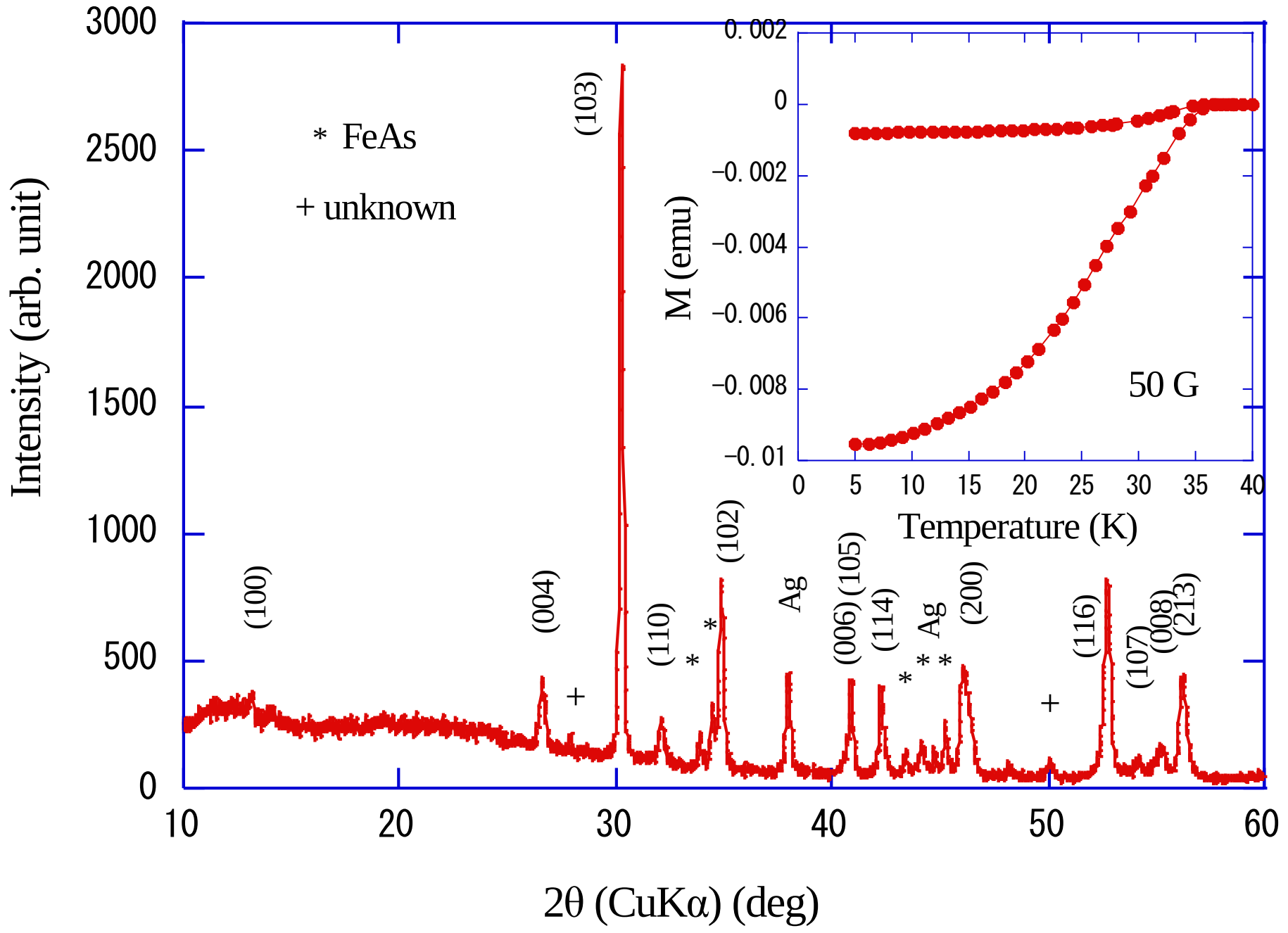

Fig. 4

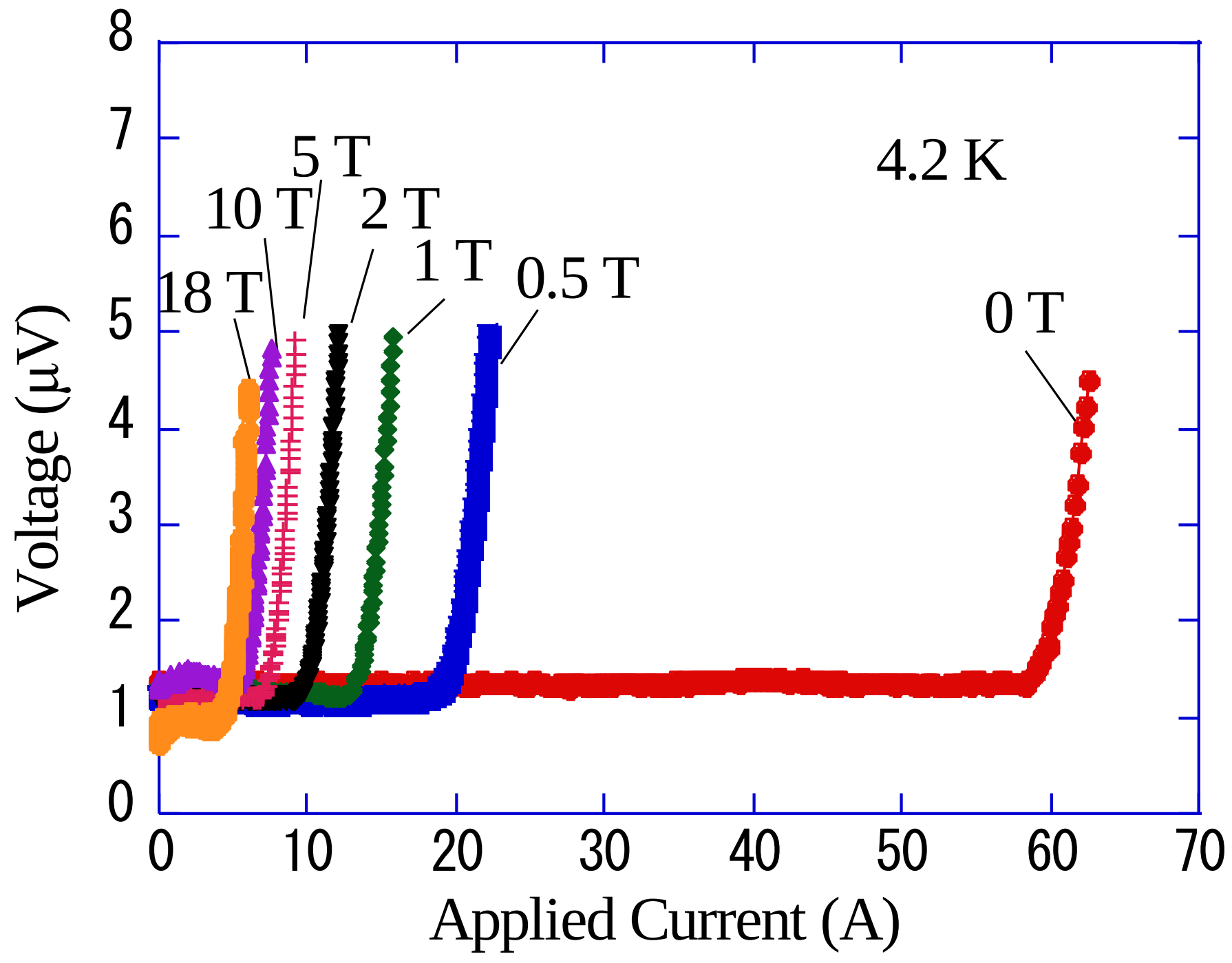

Fig. 5

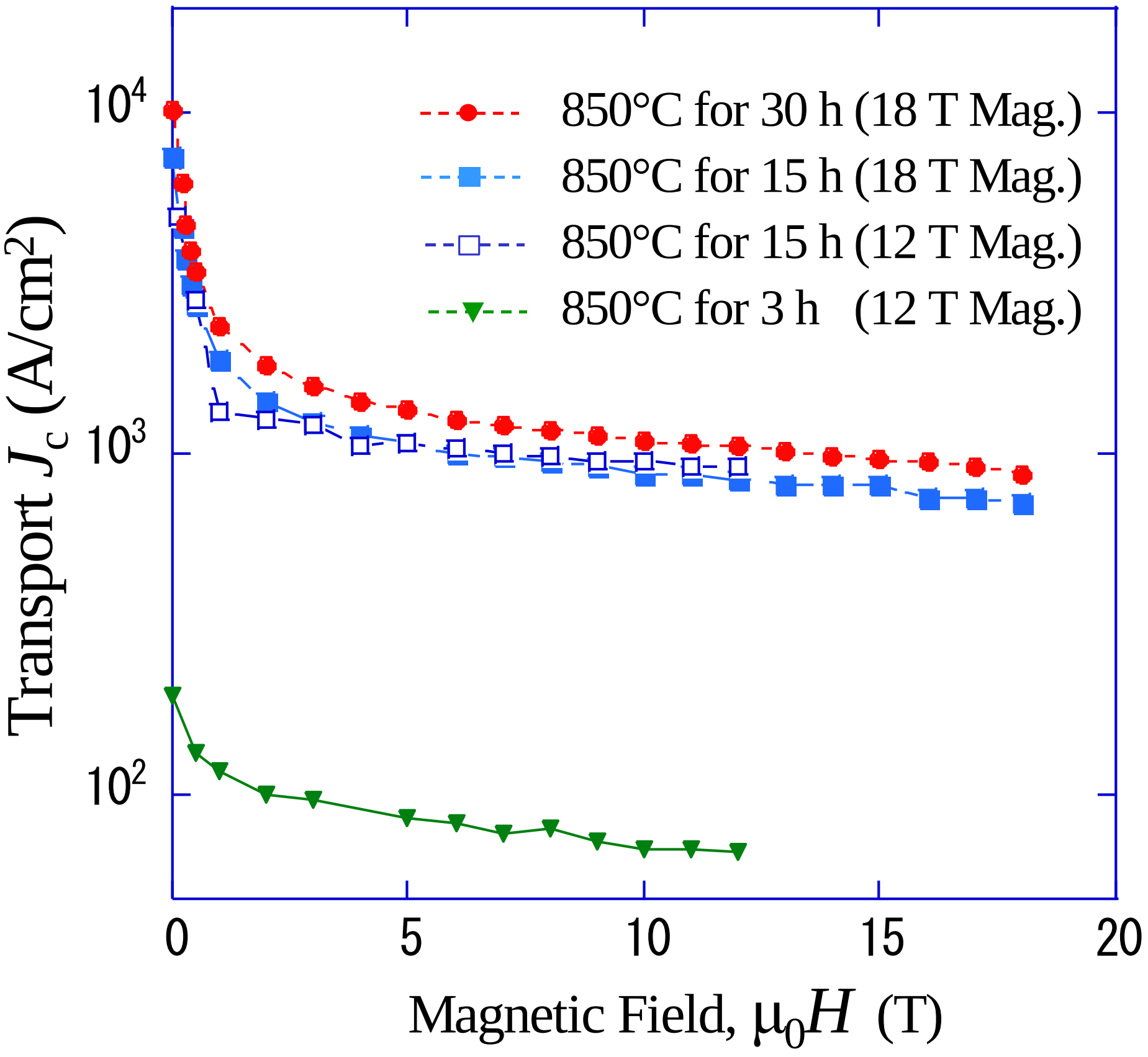

$10^4$
$10^3$
$10^2$
Transport $J_c$ (A/cm$^2$)
850°C for 30 h (18 T Mag.)
850°C for 15 h (18 T Mag.)
850°C for 15 h (12 T Mag.)
850°C for 3 h (12 T Mag.)
0
5
10
15
20
Magnetic Field, $\mu_0 H$ (T)